\documentclass[twocolumn,showpacs,aps,epsfig,nofootinbib]{revtex4}

\usepackage{graphicx}
\usepackage{epstopdf}
\usepackage{latexsym}
\usepackage{amssymb}

\usepackage[center]{subfigure}

\begin{document}

 \newcommand{\bq}{\begin{equation}}
 \newcommand{\eq}{\end{equation}}
 \newcommand{\bqn}{\begin{eqnarray}}
 \newcommand{\eqn}{\end{eqnarray}}
 \newcommand{\nb}{\nonumber}
 \newcommand{\lb}{\label}
\newcommand{\PRL}{Phys. Rev. Lett.}
\newcommand{\PL}{Phys. Lett.}
\newcommand{\PR}{Phys. Rev.}
\newcommand{\CQG}{Class. Quantum Grav.}

\title{On strong coupling in  nonrelativistic  general covariant theory of gravity} 

\author{Kai Lin ${}^{a,b}$}
\email{k_lin@baylor.edu}

\author{Anzhong Wang ${}^{a,b}$}
\email{anzhong_wang@baylor.edu}

\author{Qiang Wu ${}^{a}$}
\email{wuq@zjut.edu.cn}

\author{Tao Zhu${}^{a}$}
\email{zhut05@gmail.com}

\affiliation{${}^{a}$ Institute  for Advanced Physics $\&$ Mathematics,   Zhejiang University of
Technology, Hangzhou 310032,  China \\
${}^{b}$ GCAP-CASPER, Physics Department, Baylor
University, Waco, TX 76798-7316, USA  
}

\date{\today}

\begin{abstract}

We study the strong coupling problem in the  Horava-Melby-Thompson setup of the Horava-Lifshitz  gravity with 
an arbitrary coupling constant $\lambda$, generalized recently by da Silva, where $\lambda$ describes the deviation of 
the theory in the infrared from general relativity that has $\lambda_{GR} = 1$. We find that a scalar field in the Minkowski 
background becomes strong coupling for processes with energy higher than $\Lambda_{\omega} [\equiv (M_{pl}/c_1)^{3/2}
M_{pl}|\lambda - 1|^{5/4}]$, where  generically $c_1 \ll M_{pl}$.  However, this problem can be cured by introducing 
a new energy scale $M_{*}$, so that $M_{*} < \Lambda_{\omega}$, where $M_{*}$ denotes the suppression energy
of  high order derivative terms of the theory.

\end{abstract}


\maketitle

\section{Introduction}
\renewcommand{\theequation}{1.\arabic{equation}} \setcounter{equation}{0}

Horava recently proposed a new theory of quantum gravity \cite{Horava}, based on the perspective that Lorentz symmetry may
appear as an emergent symmetry at low energies, but can be fundamentally absent at high energies. His starting point is the
anisotropic scalings  of  space and time,
\bq
\lb{1.1}
{\bf x} \rightarrow b^{-1} {\bf x}, \;\;\;  t \rightarrow b^{-z} t.
\eq
In $(3+1)$-dimensions, in order for the theory to be power-counting
renormalizable,   the critical exponent $z$  needs to be $z \ge 3$ \cite{Visser}.
At long distances, all the high-order curvature terms are negligible, and the linear term $R$ becomes dominant. Then,
the theory is expected to  flow to the relativistic  fixed point $z = 1$, whereby the general covariance is ``accidentally restored."
The special role of time  is  realized with the Arnowitt-Deser-Misner   decomposition  \cite{ADM},
 \bqn
 \lb{1.2}
ds^{2} &=& - N^{2}c^{2}dt^{2} + g_{ij}\left(dx^{i} + N^{i}dt\right)
     \left(dx^{j} + N^{j}dt\right), \nb\\
     & & ~~~~~~~~~~~~~~~~~~~~~~~~~~~~~~  (i, \; j = 1, 2, 3).~~~
 \eqn
 Under the rescaling (\ref{1.1})
 with   $z = 3$,     $N, \; N^{i}$ and $g_{ij}$ scale, 
respectively,  as,
 \bq
 \lb{1.3}
  N \rightarrow  N ,\;  N^{i}
\rightarrow b^{-2}N^{i},\; g_{ij} \rightarrow g_{ij}.
 \eq

 The gauge symmetry of the system now is broken down  to the  foliation-preserving
diffeomorphisms Diff($M, \; {\cal{F}}$),
\bq
\lb{1.4}
\delta{t} = - f(t),\; \;\; \delta{x}^{i}  =   - \zeta^{i}(t, {\bf x}),
\eq
under which, $N,\; N^{i}$ and $g_{ij}$ transform as,
\bqn
\lb{1.5}
\delta{N} &=& \zeta^{k}\nabla_{k}N + \dot{N}f + N\dot{f},\nb\\
\delta{N}_{i} &=& N_{k}\nabla_{i}\zeta^{k} + \zeta^{k}\nabla_{k}N_{i}  + g_{ik}\dot{\zeta}^{k}
+ \dot{N}_{i}f + N_{i}\dot{f}, \nb\\
\delta{g}_{ij} &=& \nabla_{i}\zeta_{j} + \nabla_{j}\zeta_{i} + f\dot{g}_{ij},
\eqn
where $\dot{f} \equiv df/dt$,  $\nabla_{i}$ denotes the covariant
derivative with respect to the 3-metric $g_{ij}$,   $N_{i} = g_{ik}N^{k}$, and
$\delta{g}_{ij} = \tilde{g}_{ij}(t, x^{k}) - g_{ij}(t, x^{k})$, etc.  Eq.(\ref{1.5}) shows clearly  that the
lapse function $N$ and the shift vector $N^{i}$ play the role of gauge fields of the Diff($M, \; {\cal{F}}$). 
Therefore, it is natural to assume that $N$ and $N^{i}$ inherit the same dependence on
space and time as the corresponding generators,
\bq
\lb{1.6}
N = N(t), \;\;\; N_{i} = N_{i}(t, x),
\eq
while    the dynamical variables $g_{ij}$ in general   depend on both time and space, 
$g_{ij} = g_{ij}(t, x)$.
This is  often  referred to as the {\em projectability condition}.

Abandoning  the general covariance, on the other hand,   gives rise  to   a proliferation of independently  coupling constants,
which  could potentially limit   the prediction powers of the theory. Inspired by condensed matter systems \cite{Cardy},
Horava assumed that the gravitational potential ${\cal{L}}_{V}$ can be obtained from a superpotential $W_{g}$ via
the relations,
\bqn
\lb{1.7}
{\cal{L}}_{V, detailed} &=& w^{2} E_{ij}{\cal{G}}^{ijkl}E_{kl},\nb\\
E^{ij} &=& \frac{1}{\sqrt{g}}\frac{\delta{W}_{g}}{\delta{g}_{ij}},
\eqn
where $w$ is a coupling constant, and ${\cal{G}}^{ijkl}$ denotes the generalized De Witt metric, defined as
${\cal{G}}^{ijkl} = \big(g^{ik}g^{jl} + g^{il}g^{jk}\big)/2 - \lambda g^{ij}g^{kl}$, with $\lambda$ being a coupling constant.
The general covariance, $\delta{x}^{\mu} = \zeta^{\mu}(t, x), \; (\mu = 0, 1, ..., 3)$, requires $\lambda = 1$. 
The superpotential $W_{g}$ is given by
\bq
\lb{1.9}
W_{g}  = \int_{\Sigma}{\omega_{3}(\Gamma)} +  \frac{1}{\kappa^{2}_{W}}\int{d^{3}x\sqrt{g}(R - 2\Lambda)},
\eq
with $\omega_{3}(\Gamma)$ being the gravitational Chern-Simons term,
\bq
\lb{1.10}
\omega_{3}(\Gamma) = {\mbox{Tr}} \Big(\Gamma \wedge d\Gamma + \frac{2}{3}\Gamma\wedge \Gamma\wedge \Gamma\Big).
\eq
 The condition (\ref{1.7}) is  usually referred to as the {\em  detailed balance condition}.
 
However, with this condition  it was found that 
 the Newtonian  limit does not exist \cite{LMP}, and  a scalar field in the UV is not stable  \cite{CalA}. Thus, it is
 generally believed that this condition should be abandoned  \cite{KK}.  But, it has   several remarkable features \cite{Hreview}:  it is in the same spirit of the
AdS/CFT correspondence \cite{AdSCFT}; and in the nonequilibrium  thermodynamics,  the counterpart of the superpotential $W_{g}$
plays the role of entropy, while the term $E^{ij}$ the entropic forces  \cite{OM}. This might shed   light on the nature of the gravitational forces, as proposed recently
by Verlinde \cite{Verlinde}.  Due to    these desired properties,  together with Borzou,  two of the present authors recently studied this condition in detail, and
found that   the scalar field  can be stabilized, if the detailed balance  condition is allowed to be softly broken \cite{BLW}. This can also solve  the other problems
  \cite{LMP,KK}. In addition, such a  breaking can still reduce significantly the number of independent coupling constants.
For detail, we refer readers to \cite{BLW}.

It should be noted that, even  the detailed balance condition is allowed to be broken softly, the theory  is  still plagued with several other problems,
including  the instability,  ghost,  and strong coupling \cite{HWW,Mukc,Sotiriou,Padilla,GLLSW}.  To overcome those problems,   recently Horava and Melby-Thompson (HMT) \cite{HMT}   extended
the foliation-preserving-diffeomorphisms Diff($M, \; {\cal{F}}$) to include  a local $U(1)$ symmetry,
\bq
\lb{symmetry}
 U(1) \ltimes {\mbox{Diff}}(M, \; {\cal{F}}).
 \eq
 Such an extended symmetry is realized by introducing a $U(1)$ gauge field $A$ and a Newtonian prepotential $\varphi$.
 Under  Diff($M, \; {\cal{F}}$), these fields transform as  \cite{HMT,WW,HW},
\bqn
\lb{2.2}
\delta{A} &=& \zeta^{i}\partial_{i}A + \dot{f}A  + f\dot{A},\nb\\
\delta \varphi &=&  f \dot{\varphi} + \zeta^{i}\partial_{i}\varphi,
\eqn
while under  $U(1)$, characterized by the generator $\alpha$, 
they, together with $N,\; N^{i}$ and $g_{ij}$,
 transform as
\bqn
\lb{2.3}
\delta_{\alpha}A &=&\dot{\alpha} - N^{i}\nabla_{i}\alpha,\;\;\;
\delta_{\alpha}\varphi = - \alpha,\nb\\
\delta_{\alpha}N_{i} &=& N\nabla_{i}\alpha,\;\;\;
\delta_{\alpha}g_{ij} = 0 = \delta_{\alpha}{N}.
\eqn
HMT  showed   that, similar to GR, the spin-0 graviton is eliminated \cite{HMT}. This was further confirmed in \cite{WW}. 
Then, the instability of the spin-0 gravity is out of question.  In addition, in the linearized theory  Horava noticed that the U(1) symmetry only pertains to the case
$\lambda = 1$ \cite{Horava}, and it was believed that this was also the case when the Newtonian prepotential is introduced \cite{HMT}. If this were true, the ghost and
strong coupling problems would be also resolved, as both of them are related to the very fact that $\lambda \not= 1$ \cite{HW}.

However,  it has been soon challenged by da Silva \cite{Silva}, who argued  that the introduction of the   Newtonian pre-potential is so strong
that actions with $\lambda \not=1$ also has the   extended symmetry,  Eq.(\ref{symmetry}). Although   the spin-0 graviton is 
also eliminated even with any $\lambda$ as shown explicitly in  \cite{Silva,HW,Kluson},    the  ghost and strong coupling problems
 rise again, because now   $\lambda$ can be different from one.   Indeed, it was shown  \cite{HW} that to avoid the ghost problem,   $\lambda$
must  satisfy  the same constraints,
\bq
\lb{1.11}
\lambda \ge 1,\;\;\; {\mbox{or }} \;\;\;  \lambda < \frac{1}{3},
\eq
as found previously \cite{Horava,SVW,WM,BS}. 

In addition,   the strong coupling problem also arise  \cite{HW}. In this paper, we shall address this important issue. In particular,  in Sec. II we  briefly review the HMT setup with any coupling
constant $\lambda$
with detailed balance condition softly breaking,  a version presented in \cite{BLW} in detail, while in Sec. III we study the strong coupling problem when a scalar field is present. We find that
  a scalar field in the Minkowski background becomes strong coupling for processes with energy higher than $\Lambda_{\omega} \equiv (M_{pl}/c_1)^{3/2} M_{pl}|\lambda - 1|^{5/4}$. For
$c_1 \simeq M_{pl}$, this gives precisely the strong coupling strength found in \cite{HW}. However, this  problem can be resolved  by introducing a new energy scale 
$M_{*}$ \cite{BPS}, so that $M_{*} \lesssim \Lambda_{\omega}$, where  $M_{*}$ defines the  energy scale that suppresses  the sixth-order 
derivative terms of the theory. In Sec. IV, we present our main conclusions. 

It should be noted that the strong coupling problem in other versions of the Horava-Lifshitz (HL) theory has been studied extensively by using both effective field theory \cite{WWb,HW,KA,PS,CPS} and 
St\"uckelberg formalism \cite{BPSb,BPS,KP}. In this paper, we shall follow the approach of the effective field theory \cite{Pol}, although the final conclusions  are independent of the methods
to be used. In addition, strong coupling can happen not only due to gravitational/matter self-interactions, but also  to the interactions among gravitational and matter fields. The latter
was studied in \cite{CPS,HW}, while strong coupling due to the self-gravitational interactions were studied in \cite{BPSb,KA,PS,BPS,WWb}. In this paper, we shall study it due to the interaction 
between   gravitational and  scalar fields.

\section{General covariant theory with detailed balance condition softly breaking}

\renewcommand{\theequation}{2.\arabic{equation}} \setcounter{equation}{0}

As mentioned above,   HMT considered only the case  $\lambda = 1$. Later, da Silva
generalized it to the cases with any $\lambda$ \cite{Silva}, in which the total action  can be written in the form  \cite{Silva,HW},
 \bqn \lb{2.4}
S &=& \zeta^2\int dt d^{3}x N \sqrt{g} \Big({\cal{L}}_{K} -
{\cal{L}}_{{V}} +  {\cal{L}}_{{\varphi}} +  {\cal{L}}_{{A}} +  {\cal{L}}_{{\lambda}} \nb\\
& & ~~~~~~~~~~~~~~~~~~~~~~ \left. +{\zeta^{-2}} {\cal{L}}_{M} \right),
 \eqn
where $g={\rm det}\,g_{ij}$,
 \bqn \lb{2.5}
{\cal{L}}_{K} &=& K_{ij}K^{ij} -   \lambda K^{2},\nb\\
{\cal{L}}_{\varphi} &=&\varphi {\cal{G}}^{ij} \Big(2K_{ij} + \nabla_{i}\nabla_{j}\varphi\Big),\nb\\
{\cal{L}}_{A} &=&\frac{A}{N}\Big(2\Lambda_{g} - R\Big),\nb\\
{\cal{L}}_{\lambda} &=& \big(1-\lambda\big)\Big[\big(\Delta\varphi\big)^{2} + 2 K \Delta\varphi\Big],
 \eqn
 $\Delta \equiv \nabla^{2} = g^{ij}\nabla_{i}\nabla_{j}$, and 
$\Lambda_{g}$ is a    coupling constant. The
Ricci and Riemann terms all refer to the 3-metric $g_{ij}$, and
 \bqn
  \lb{2.6}
K_{ij} &=& \frac{1}{2N}\left(- \dot{g}_{ij} + \nabla_{i}N_{j} + \nabla_{j}N_{i}\right),\nb\\
{\cal{G}}_{ij} &=& R_{ij} - \frac{1}{2}g_{ij}R + \Lambda_{g} g_{ij}.
 \eqn

${\cal{L}}_{{V}}$ is an arbitrary Diff($\Sigma$)-invariant local scalar functional
built out of the spatial metric, its Riemann tensor and spatial covariant derivatives. With the detailed balance condition softly breaking,
it takes the form \cite{BLW,KKb},
 \bqn \lb{2.5a}
{\cal{L}}_{{V}} &=& \zeta^{2}\gamma_{0}  + \gamma_{1} R + \frac{1}{\zeta^{2}}
\Big(\gamma_{2}R^{2} +  \gamma_{3}  R_{ij}R^{ij}\Big)\nb\\
& &  ~~~ +  \frac{\gamma_{4}}{\zeta^{3}} \epsilon^{ijk}R_{il}\nabla_{j}R^{l}_{k} +  \frac{\gamma_{5}}{\zeta^{4}} C_{ij}C^{ij}, ~~~
 \eqn
 where the coupling  constants $ \gamma_{s}\, (s=0, 1, 2,\dots 5)$  are all dimensionless, and  ${\gamma_{5}} \equiv w^{2}{\zeta^{4}}$. The  relativistic  limit in the IR
 requires
 \bq
 \lb{gamma}
  \lambda = 1,\;\;\;
  \gamma_{1} = -1,\;\;\;
 \zeta^{2} = \frac{1}{16\pi G}.
 \eq
 The existence of the $\gamma_{4}$ term explicitly breaks the
 parity,  which could have  important observational consequences on primordial gravitational waves \cite{TS}.

 ${\cal{L}}_{M}$ is the
matter Lagrangian density.  For a scalar field $\chi$ with detailed balance conditions softly breaking, it  is given by  \cite{BLW},
\bq
\lb{2.6a}
{\cal{L}}_{M} =  {\cal{L}}^{(A,\varphi)}_{\chi} + {\cal{L}}^{(0)}_{\chi},
\eq
where
\bqn
\lb{2.7}
{\cal{L}}^{(A,\varphi)}_{\chi} &=& \frac{A - {\cal{A}}}{N}  \Big[c_{1}\Delta\chi + c_{2}\big(\nabla\chi\big)^{2}\Big]\nb\\
&&  - \frac{f}{N}\Big(\dot{\chi} - N^{i}\nabla_{i}\chi\Big)\big(\nabla^{k}\varphi\big)\big( \nabla_{k}\chi\big)\nb\\
& & + \frac{f}{2}\Big[\big(\nabla^{k}\varphi\big)\big(\nabla_{k}\chi\big)\Big]^{2},\nb\\
{\cal{L}}^{(0)}_{\chi} &=& \frac{f}{2N^{2}}\Big(\dot{\chi} - N^{i}\nabla_{i}\chi\Big)^{2} - {\cal{V}},
\eqn
and
\bqn
\lb{2.8}
{\cal V}  &=& V\left(\chi\right) + \left({1\over2}+V_{1}\left(\chi\right)\right) (\nabla\chi)^2
+  V_{2}\left(\chi\right){\cal{P}}_{1}^{2}\nb\\
& & +  V_{3}\left(\chi\right){\cal{P}}_{1}^{3}  +
V_{4}\left(\chi\right){\cal{P}}_{2} + V_{5}\left(\chi\right)(\nabla\chi)^2{\cal{P}}_{2}\nb\\
& & + V_{6} {{\cal P}}_{1} {\cal{P}}_{2},
\eqn
with $V(\chi)$ and $V_{n}(\chi)$ being arbitrary functions of $\chi$, and
\bq
\lb{2.9}
{{\cal P}}_{n} \equiv    \Delta^{n}\chi, \;\;\;
V_{6} \equiv - \sigma^{2}_{3}.
\eq

The corresponding field equations are given in Appendix A and by Eqs.(3.13) and (3.14) in  \cite{BLW}.

\section{Strong Coupling}

\renewcommand{\theequation}{3.\arabic{equation}} \setcounter{equation}{0}

To study the strong coupling problem, it is found sufficient to consider perturbations of the Minkowski spacetime,
\bqn
\lb{3.1}
&& \bar{N} = 1,\;\;\; \bar{N}_{i} = 0,\;\;\; \bar{g}_{ij} = \delta_{ij},\;\;\;
\bar{A} = \bar{\varphi} = 0,\nb\\
&& \bar{\chi} = \bar{\chi}_{0}, \;\;\; V(\bar{\chi}_{0}) = V'(\bar{\chi}_{0}) = 0 = \Lambda = \Lambda_{g},
\eqn
where   $\chi_{0}$ is a constant. Without loss of generality, we can set  $\chi_{0} = 0$, a condition that will be assumed in the rest of the paper.
The perturbed fields with the (generalized)  quasilongitudinal gauge are given by \cite{WM,HW,WW},
\bqn
\lb{3.2}
&& N = 1,  \;\;\; N_{i} = B_{,i}, \;\;\;\; g_{ij} = (1-2\psi)\delta_{ij},\nb\\
&& A = \delta{A},\;\;\; \varphi = 0, \;\;\; \chi = \delta\chi.
\eqn
Then,   we find
\bqn
\lb{3.3}
N^{i} &=& \big(1 + 2\psi + 4\psi^{2}\big)B^{,i} + {\cal{O}}(\epsilon^{4}),\nb\\
g^{ij} &=& \big(1 + 2\psi + 4\psi^{2} + 8\psi^{3}\big)\delta^{ij} + {\cal{O}}(\epsilon^{4}),\nb\\
\sqrt{g} &=& 1- 3\psi + \frac{3}{2}\psi^{2} + \frac{1}{2}\psi^{3} + {\cal{O}}(\epsilon^{4}),
\eqn
where $B^{,i} \equiv \delta^{ij}B_{,j}$, etc. After simple but  tedious calculations, to third order we find that
${\cal{L}}_{\lambda} = {\cal{L}}_{\varphi} = 0$, and
\bqn
\lb{3.4}
\sqrt{g}{\cal{L}}_{K} &=& (1+ \psi)\Big[(1-3\lambda)\big(3\dot{\psi}^{2} + 2\dot{\psi}\partial^{2}B\big) + B_{,ij}B^{;ij} \nb\\
& & ~~~~~~~~~~~ - \lambda\big(\partial^{2}B\big)^{2}\Big] + 2\Big[2B_{,ij}B^{,i}\psi^{,j} \nb\\
& & - \big(1-3\lambda\big)\dot{\psi}B_{,i}\psi^{,i}
- \big(1-\lambda\big)B_{,i}\psi^{,i}\partial^{2}B\Big], \nb\\
\sqrt{g}{\cal{L}}_{A} &=& - 2{A}\Big[2\big(1 + \psi\big)\partial^{2}\psi + 3\psi_{,i}\psi^{,i}\Big],\nb\\
\sqrt{g}{\cal{L}}_{V} &=& 4\partial^{2}\psi +  {}^{(2)}\Big(\sqrt{g}{\cal{L}}_{V}\Big) +  {}^{(3)}\Big(\sqrt{g}{\cal{L}}_{V}\Big),\nb\\
\sqrt{g}{\cal{L}}_{\chi} &=& {}^{(2)}\Big(\sqrt{g}{\cal{L}}_{\chi}\Big) +  {}^{(3)}\Big(\sqrt{g}{\cal{L}}_{\chi}\Big),
\eqn
where the first order term $\partial^{2}\psi$ in the expression of $(\sqrt{g}{\cal{L}}_{V})$ becomes a boundary term
after integration, and therefore can be discarded, and
\bqn
\lb{3.5a}
{}^{(2)}\Big(\sqrt{g}{\cal{L}}_{V}\Big) &=& 2\psi\partial^{2}\psi 
+ \frac{2}{\zeta^{2}}\big(8\gamma_{2}+3\gamma_{3}\big) \psi\partial^{4}\psi,\nb\\
 {}^{(2)}\Big(\sqrt{g}{\cal{L}}_{\chi}\Big)  &=&\frac{f}{2}{\dot\chi}^{2} - \frac{1}{2}V''\chi^{2}
 + \frac{1}{2}\big(1 +2V_{1}\big)\chi\partial^{2}\chi\nb\\
 & &
 - \big(V_{2} + V_{4}'\big)\chi\partial^{4}\chi + \sigma^{2}_{3}\chi\partial^{6}\chi\nb\\
 & & + c_{1}\chi\partial^{2}{A},\\
\lb{3.5b}
{}^{(3)}\Big(\sqrt{g}{\cal{L}}_{V}\Big) &=&- 6\big(\psi^2\partial^{2}\psi + 3 \psi\psi_{,i}\psi^{,i}\big) \nb\\
&& + A_{1}(\gamma_2, \gamma_3)\psi^{2} \partial^{4}\psi + {\cal{O}}(\gamma_2, \gamma_3, \psi, \partial^{4})\nb\\
& & + A_{2}(\gamma_5)\psi^{2}\partial^{6} \psi +  {\cal{O}}(\gamma_5,  \psi, \partial^{6}),\nb\\
 {}^{(3)}\Big(\sqrt{g}{\cal{L}}_{\chi}\Big)  &=& - \frac{3}{2}f\psi\dot{\chi}^{2}
 -f \dot{\chi}B_{,i}\chi^{,i}  -c_1 \psi {A} \partial^{2}\chi\nb\\
 & & - c_1{A} \psi^{,i}\chi_{,i} + c_2{A} \chi^{,i}\chi_{,i} + \frac{3}{2}V''\psi\chi^{2}\nb\\
 &&
 - \frac{1}{6}V'''\chi^{3} - V_{1}'\chi \chi_{,i}  \chi^{,i} \nb\\
 & & + \frac{1}{2}\big(1 + 2V_1\big)\psi {\chi}_{,i} \chi^{,i}\nb\\
&& + B_{1}\big(V_2, V_4\big)\chi^{2} \partial^{4}\chi\nb\\
& & ~~~~~~ + {\cal{O}}(V_2, V_4, \chi, \partial^{4})\nb\\
& & + B_{2}\big(V_3, V_5, V_6\big)\chi^{2} \partial^{6}\chi \nb\\
& & ~~~~~~  +  {\cal{O}}(V_3, V_5, V_6, \chi, \partial^{6}),
\eqn
where the terms of $A_{i}$ ($B_{i}$) are representatives of the fourth and sixth order derivative terms of $\psi$ ($\chi$), and the specific
dependences of them  on their arguments are not important to the analysis of the strong coupling problem, as shown below. So, for the sake
of simplicity,  we shall not give them explicitly here.  Hence, to second order we obtain
\bqn
\lb{3.6}
S^{(2)} &=&\zeta^{2} \int{dt d^{3}x}\Bigg\{3(1-3\lambda)\dot{\psi}^{2} + 2 (1-3\lambda)\dot{\psi}\partial^{2}B \nb\\
& & + (1-\lambda)\big(\partial^{2}B\big)^{2}
- 4{\psi}\partial^{2}{A} - 2 {\psi}\partial^{2}\psi\nb\\
&&  - \frac{2}{\zeta^{2}}(8\gamma_2 + 3\gamma_3)\psi\partial^{4}\psi \nb\\
& & + \frac{1}{\zeta^{2}}\Bigg[\frac{1}{2}f{\dot\chi}^{2} - \frac{1}{2}V''\chi^{2}
 + \frac{1}{2}\big(1 +2V_{1}\big)\chi\partial^{2}\chi\nb\\
 & & ~~~~~~~~~
 - \big(V_{2} + V_{4}'\big)\chi\partial^{4}\chi + \sigma^{2}_{3}\chi\partial^{6}\chi\nb\\
 & &  ~~~~~~~~~ + c_{1}\chi\partial^{2}{A}\Bigg]\Bigg\}.
\eqn
Variations of $S^{(2)}$, respectively, with respect to $\psi,\; B,\; {A}$ and $\chi$ yield,
\bqn
\lb{3.7a}
&& \ddot{\psi} + \frac{1}{3}\partial^{2}\dot{B} = \frac{2}{3(3\lambda - 1)}\partial^{2}\Bigg({A}  + \psi\nb\\
&& ~~~~~~~~~~~~~~~~  + \frac{8\gamma_2 + 3\gamma_3}{\zeta^{2}}\partial^{2}\psi\Bigg),\\
\lb{3.7b}
&& (1-3\lambda)\dot{\psi} + (1-\lambda)\partial^{2}B = 0,\\
\lb{3.7c}
&& \psi = 4\pi G c_{1}\chi,
\eqn
and
\bqn
\lb{3.8}
&& f\ddot{\chi} + V''\chi - (1 + 2V_1)\partial^{2}\chi + 2(V_2 + V_4')\partial^{4}\chi\nb\\
&& ~~~~~~~~~~~~~~~~~~~~~~~~~~~~ - 2\sigma^{2}_{3}\partial^{6}\chi = c_1\partial^{2}{A}.
\eqn
The above equations can be obtained from   Eqs.(4.13) - (4.17) and Eq.(4.20) of \cite{BLW}, by setting
$\bar{\chi} = \bar{\chi}' = 0$ and $a = 1$, as it is expected. Using Eqs.(\ref{3.7a})-(\ref{3.7b}), we can integrate out $\psi,\; B$ and ${A}$, so
$S^{(2)}$   finally takes the form,
\bqn
\lb{3.9}
S^{(2)} &=& \beta^{2}\int{dtd^{3}x\Big[\dot\chi^{2} - \alpha\big(\partial\chi\big)^{2} - m_{\chi}^{2} \chi^{2} }\nb\\
& & ~~~~~~~~~ -  \frac{1}{M_{A}^{2}} \chi\partial^{4}\chi
        + \frac{1}{M_{B}^{4}} \chi\partial^{6}\chi\Big], ~~~~~~~
\eqn
where
\bqn
\lb{3.10}
\beta^{2} &\equiv &\frac{2\pi G c_{1}^{2}}{|c_{\psi}|^{2}}   + \frac{f}{2},\nb\\
\alpha &\equiv& \frac{1}{2\beta^{2}}\Big(1 + 2V_{1} - 4\pi G c_{1}^{2}\Big),\nb\\
M^{2}_{A} &\equiv& \beta^{2}\Big(2\pi G c_{1}^{2}  \frac{8\gamma_2 + 3\gamma_3}{\zeta^{2}}\ + V_{2} + V_{4}'\Big)^{-1},\nb\\
M^{4}_{B} &\equiv& \frac{\beta^{2}}{\sigma_{3}^{2}},\;\;\; m^{2}_{\chi} \equiv \frac{1}{2\beta^{2}}V'',
\eqn
and $c_{\psi}^{2} = (1-\lambda)/(3\lambda - 1)$. As a consistency check, one can show that the variation of the action (\ref{3.9})  with respect to $\chi$ yields the master equation (4.21) given in \cite{BLW}.
In addition, when $\lambda$ satisfies the condition (\ref{1.11}), the above expression shows clearly that the scalar field is ghost free for $f > 0$, as
first noticed in \cite{BLW}. The scalar field is stable in all energy scales by properly choosing the potential terms $V_{i}$, including the  UV and IR.
For detail, we refer readers to \cite{BLW}. 

Therefore, in the following we focus only  on the strong coupling problem.
 To this end, let us first note that the cubic action is given by \footnote{Note  the difference between the perturbations considered in
 \cite{KA,PS,WW,HW} and the ones studied here. In particular, in   \cite{KA,PS,WW,HW} the  perturbations of the form,
 $N = 1, \; N_{i} = \partial_{i}\beta,\; g_{ij} = e^{2\zeta}\delta_{ij}$, were studied, while in this paper   we
 use the expansions of Eq.(\ref{3.3}) to calculate the third order action. Although such obtained $S^{(3)}$ is different from the one obtained in
 \cite{KA,PS,WW,HW},  it is not difficult to argue that  the final conclusions for the strong coupling
 problem will be the same.},
  \bqn
 \lb{3.10a}
 S^{(3)} &=& \int{dtd^{3}x\Bigg\{\lambda_{1}\left(\frac{1}{\partial^{2}}\ddot{\chi}\right)\chi\partial^{2}\chi  + \lambda_{2}\left(\frac{1}{\partial^{2}}\ddot{\chi}\right)\chi_{,i}\chi^{,i}}
 \nb\\
 && +  \lambda_{3} {\chi}\dot\chi^{2} + \lambda_{4} \chi_{,i}\left(\frac{\partial_{j}}{\partial^{2}}\dot\chi\right)\left(\frac{\partial^{i}\partial^{j}}{\partial^{2}}\dot\chi\right)
 \nb\\
 &&  +  \lambda_{5} \dot\chi \chi^{,i}\left(\frac{\partial_{i}}{\partial^{2}}\dot\chi\right) + \lambda_{6}\chi^{3} + \lambda_{7} \chi^{2}\partial^{2}\chi  \nb\\
 && +  \lambda_{8} \chi^{2} \partial^{4}\chi  + \lambda_{9}  \chi^{2} \partial^{6}\chi \nb\\
 & &  + ...\Bigg\},
 \eqn
 where ``..." represents the fourth- and sixth-order derivative terms of   ${\cal{O}}$'s given in Eq.(\ref{3.5b}), which are irrelevant to the strong coupling
 problem, as mentioned above, and 
 \bqn
 \lb{3.10ba}
 \lambda_{1} &=& \frac{c_1^3}{8\zeta^4|c_\psi|^{2}},\;\;\;
 \lambda_{2} =  \frac{1}{|c_\psi|^{2}}\left(\frac{5c_1^3}{32\zeta^4}-\frac{c_1c_2}{4\zeta^2}\right),\nb\\
 \lambda_{3} &=& \frac{c_1^3}{32\zeta^4 |c_\psi|^{2}} -\frac{3fc_1}{8\zeta^2},\;\;\;
 \lambda_{4} =  \frac{3c_1^3}{64\zeta^4 |c_\psi|^{4}}, \nb\\
 \lambda_{5} &=& \frac{c_1^3}{64\zeta^4|c_\psi|^{4}} +\frac{c_1f}{4\zeta^2 |c_\psi|^{2}}, \;\;\;
 \lambda_{6} = \frac{3c_1}{8\zeta^2}\ddot{V}-\frac{\dddot{V}}{6}, \nb\\
 \lambda_{7} &=&
 \frac{\dot{V}_1}{2}+\frac{c_1^2c_2}{8\zeta^2}-\frac{c_1}{16\zeta^2}-\frac{c_1V_1}{8\zeta^2}, \nb\\
 \lambda_{8} &=& \tilde{A}_{1}(\gamma_2, \gamma_3, c_1)     + {B}_{1}\big(V_2, V_4\big), \nb\\
  \lambda_{9} &=& \tilde{A}_{2}(\gamma_5, c_1)  + {B}_{2}\big(V_3, V_5, V_6\big),
 \eqn
 where  $\tilde{A}_{i} \equiv (4\pi G c_{1})^{3}A_{i}$.  
 Depending on the energy scales, each term in Eq.(\ref{3.10a}) will have different scalings. Thus, 
  in the following  we consider them separately.

 \subsection{$|\nabla| \ll M_{*}$}

When $|\nabla| \ll M_{*}$, where $M_{*} = {\mbox{Min.}}\big(M_{A}, M_{B}\big)$,  
 we find that the high order derivative terms in Eq.(\ref{3.9}) can be neglected, and 
 \bq
\lb{3.11} S^{(2)} \simeq \beta^{2}\int{dtd^{3}x\Big[\dot\chi^{2} -
\alpha\big(\partial\chi\big)^{2}\Big]}. 
\eq
Note that in writing the above expression, without loss of generality, we had assumed that $|\nabla| \gg m_{\chi}$.
Setting
\bq 
\lb{3.12} 
t = b_{1}\hat{t},\;\;\; x^{i} = b_{2}\hat{x}^{i},
\;\;\; \chi= b_{3}\hat\chi, 
\eq 
we can bring Eq.(\ref{3.11}) into its ``canonical" form,
\bq 
\lb{3.13} 
S^{(2)} \simeq
\int{d\hat{t}d^{3}\hat{x}\Big[\big(\hat\chi^{*}\big)^{2} -
\big(\hat{\partial}\hat{\chi}\big)^{2}\Big]}, 
\eq
in which the coefficient of each term is order of  one, for
\bq 
\lb{3.14}
b_{2} = b_{1}\sqrt{\alpha},\;\;\; b_{3} =\frac{1}{b_{1}\beta\alpha^{3/4}},
\eq
where $\hat\chi^{*} \equiv d\hat\chi/d\hat{t}$. Note that the requirement that the coefficient of each term be  order of one is 
important in order to obtain a correct coupling strength \cite{BPS,WWb,HW}. In addition, the transformations (\ref{3.12})
should not be confused with the gauge choice (\ref{3.2}), as it just provides a technique to obtain the correct  coupling strength.
In fact, when we consider physics, we will all refer to the ones obtained in the $t$ and $x$ coordinates, 
as to be shown below.  
Inserting Eq.(\ref{3.12})   into Eq.(\ref{3.10a}), we obtain
  \bq
 \lb{3.10bb}
 S^{(3)} = \frac{1}{b_{1}\beta^{3}\alpha^{3/4}} \hat{S}^{(3)},
 \eq
where
\bqn
\lb{3.10b} 
  \hat{S}^{(3)} &\equiv&
 \int{d\hat{t}d^{3}\hat{x}\Bigg\{\lambda_{1}\left(\frac{1}{\hat\partial^{2}}\hat{\chi}^{**}\right)\hat\chi\hat\partial^{2}\hat\chi}\nb\\
 && + \lambda_{2}\left(\frac{1}{\hat\partial^{2}}\hat{\chi}^{**}\right)\hat\partial_{i}\hat\chi\hat\partial^{i}\hat\chi
 +  \lambda_{3} {\hat\chi}\hat\chi^{*2}
 \nb\\
 &&  + \lambda_{4}\left(\hat\partial_{i}\hat\chi\right)\left(\frac{\hat\partial_{j}}{\hat\partial^{2}}\hat\chi^{*}\right)\left(\frac{\hat\partial^{i}\hat\partial^{j}}{\hat\partial^{2}}\hat\chi^{*}\right)
 \nb\\
 &&  +  \lambda_{5} \hat\chi^{*} \left(\frac{\hat\partial_{i}}{\hat\partial^{2}}\hat\chi^{*}\right)\hat\partial_{i}\hat\chi\nb\\
 &&  +\lambda_{6} b_{1}^{2}\hat\chi^{3} + \frac{\lambda_{7}b_{1}^{2}}{b_{2}^{2}} \hat\chi^{2}\hat\partial^{2}\hat\chi  \nb\\
 && +   \frac{\lambda_{8}b_{1}^{2}}{b_{2}^{4}} \hat\chi^{2} \hat\partial^{4}\hat\chi  +  \frac{\lambda_{9}b_{1}^{2}}{b_{2}^{6}} \hat\chi^{2} \hat\partial^{6}\hat\chi \nb\\
 & &  + ...\Bigg\}.
 \eqn

On the other hand, from Eq.(\ref{3.13}) one finds  that $S^{(2)}$ is invariant under  the rescaling,
\bq
\lb{3.15}
\hat{t} \rightarrow b^{-1}\hat{t}, \;\;\; \hat{x}^{i} \rightarrow b^{-1}\hat{x}^{i},\;\;\; \hat{\chi} \rightarrow b\hat{\chi},
\eq
while the  terms of $\lambda_{1, 2, ..., 5}$ and $\lambda_{7}$  in $S^{(3)}$ all scale as $b$, and the terms of $\lambda_{6, 8, 9}$ scale as $b^{-1},\; b^{3}, \; b^{5}$, respectively.
Therefore, except for the $\lambda_{6}$ term, all the others are irrelevant and nonrenormalizable \cite{Pol}. For example, considering a process with  an energy $E$, then we
find that the fourth term has the contribution, 
\bq
\lb{3.16}
\int{d\hat{t}d^{3} \hat{x}\left(\hat\partial_{i}\hat\chi\right)\left(\frac{\hat\partial_{j}}{\hat\partial^{2}}\hat\chi^{*}\right)\left(\frac{\hat\partial^{i}\hat\partial^{j}}{\hat\partial^{2}}\hat\chi^{*}\right)}
\simeq E.
\eq
Since the action $S^{(3)}$ is dimensionless, we must have
\bqn
\lb{3.17}
 && \frac{\lambda_{4}}{b_{1}\beta^{3}\alpha^{3/4}} \int{d\hat{t}d^{3} \hat{x}\left(\hat\partial_{i}\hat\chi\right)\left(\frac{\hat\partial_{j}}{\hat\partial^{2}}\hat\chi^{*}\right)
 \left(\frac{\hat\partial^{i}\hat\partial^{j}}{\hat\partial^{2}}\hat\chi^{*}\right)}
\nb\\
&& ~~~~~~~~~~~~~~ \simeq \frac{E}{\Lambda^{(4)}_{SC}},
\eqn
where $\Lambda^{(4)}_{SC}$ has the same dimension of  $E$, and is given by,
\bq
\lb{3.18}
\Lambda^{(4)}_{SC} = \frac{b_{1}\beta^{3}\alpha^{3/4}}{\lambda_{4}}.
\eq
Similarly, one can find $\Lambda^{(n)}_{SC}$ for all the other nonrenormalizable terms. But,  when $\lambda \rightarrow 1$ (or $c_{\psi} \rightarrow
0$),  the lowest one of the $\Lambda^{(n)}_{SC}$'s  is given by $\Lambda^{(4)}_{SC}$, so we have
\bq
\lb{3.19}
\Lambda_{\hat{\omega}}  \equiv \frac{b_{1}\beta^{3}\alpha^{3/4}}{\lambda_{4}},
\eq
above which the nonrenormalizable $\lambda_{4}$ term becomes larger than unit, and the process runs into the strong coupling regime. 
Back to the physical coordinates  $t$ and $x$, the corresponding energy and momentum scales are given, respectively, by  
\bqn
\lb{3.20}
\Lambda_{\omega} &=& \frac{\Lambda_{\hat{\omega}}}{b_{1}} 
\simeq {\cal{O}}(1)\left(\frac{\zeta}{c_1}\right)^{3/2}M_{pl}\left|c_{\psi}\right|^{5/2},\nb\\
\Lambda_{k} &=& \frac{\Lambda_{\hat{\omega}}}{b_{2}} 
\simeq {\cal{O}}(1)\left(\frac{\zeta}{c_1}\right)^{1/2}M_{pl}\left|c_{\psi}\right|^{3/2}.
\eqn
In particular, for $c_1 \simeq \zeta$, 
we find that $\Lambda_{\omega}  \simeq M_{pl}\left|c_{\psi}\right|^{5/2}$, which is precisely the result obtained in \cite{HW}. 

\begin{figure}[tbp]
\centering
\includegraphics[width=8cm]{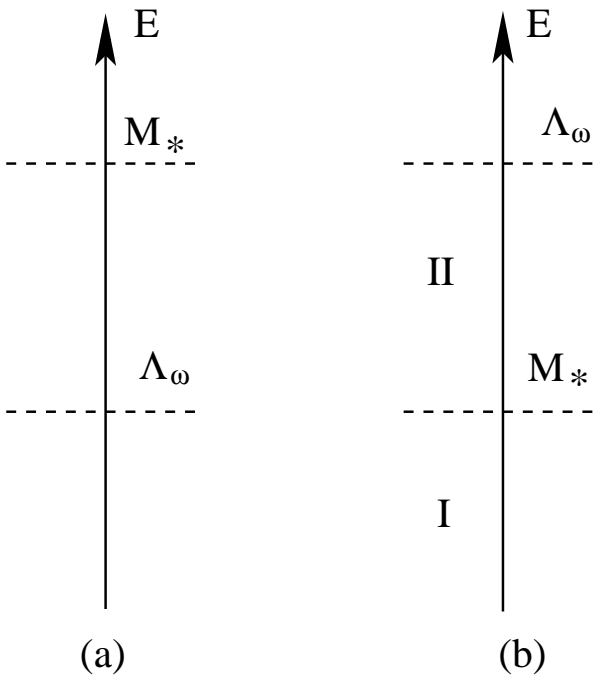}
\caption{The energy scales: (a) $\; \Lambda_{\omega} <  M_{*}$; and (b) $\; \Lambda_{\omega} > M_{*}$.}
\label{fig1}
\end{figure}

It should be noted that the above conclusion is true only for $M_{*} > \Lambda_{\omega}$, that is,
\bq
\lb{3.21}
M_{*} > \left(\frac{\zeta}{c_1}\right)^{3/2}M_{pl}\left|c_{\psi}\right|^{5/2},
\eq
as shown by Fig. \ref{fig1}(a).

When $M_{*} < \Lambda_{\omega}$, the above analysis holds only  for the processes with $E \ll M_{*}$ [Region I in Fig.\ref{fig1}(b)]. However,  
 when   $E \gtrsim M_{*}$ and before the strong coupling energy scale $\Lambda_{\omega}$
 reaches [cf. Fig.\ref{fig1}(b)], the high order derivative terms of $M_{A}$ and $M_{B}$  in Eq.(\ref{3.9}) cannot be 
 neglected any more, and one has to take these terms into account. It is exactly because the presence
 of these terms that the strong coupling problem is cured \cite{BPS}. In the following,  we show that this is also the case here in the HMT setup  \cite{HMT}.  
 
 \subsection{$M_{*} < \Lambda_{\omega}$}
 
In this case, there are two possibilities, $M_{A} < M_{B}$ and $M_{A} \gtrsim M_{B}$. In the following, let us consider them separately.

\subsubsection{$M_{A} < M_{B}$}

When $M_{A} < M_{B}$, we have $M_{*} = M_{A}$.  For the processes with $E \gtrsim M_{A}$,  Eq.(\ref{3.9}) reduces to
\bq
\lb{3.22}
S^{(2)} = \beta^{2}\int{dtd^{3}x\left(\dot\chi^{2}    -  \frac{1}{M_{A}^{2}} \chi\partial^{4}\chi\right)}.
\eq
To study the strong coupling problem, we shall follow what we did in the last case, by first  writing $S^{(2)}$ in its canonical form,
\bq
\lb{3.22a}
S^{(2)} =  \int{d\hat{t}d^{3}\hat{x}\left(\hat\chi^{*2}    - \hat\chi\hat\partial^{4}\hat\chi\right)},
\eq
 through the transformations (\ref{3.12}).  It can be shown that now $b_{2}$ and $b_{3}$ are given by
\bq
\lb{3.23}
b_{2} = \sqrt{\frac{b_{1}}{M_{A}}},\;\;\;
b_{3} = \frac{M_{A}^{3/4}}{b_{1}^{1/4}\beta},
\eq
for which the cubic action $S^{(3)}$ takes the form,
  \bq
 \lb{3.10c}
 S^{(3)} = \frac{M_{A}^{3/4}}{b_{1}^{1/4}\beta^{3}} \hat{S}^{(3)},
   \eq
where $\hat{S}^{(3)}$ is given by Eq.(\ref{3.10b}). 
Due to the nonrelativistic nature of the action (\ref{3.22a}), its scaling becomes anisotropic,
\bq
\lb{3.24}
\hat{t} \rightarrow b^{-2}\hat{t},\;\;\;
\hat{x}^{i} \rightarrow b^{-1}\hat{x}^{i},\;\;\;
\hat{\chi} \rightarrow b^{1/2}\hat{\chi}.
\eq
Then, we find that   the first five terms in Eqs.(\ref{3.10c}) and  (\ref{3.10b}) scale as $b^{1/2}$, while the terms of $\lambda_{6, ...,9}$ scale, respectively, as $b^{-7/2},\;
b^{-3/2},\; b^{1/2},\; b^{5/2}$. Thus, except for the $\lambda_{6}$ and $\lambda_{7}$ terms, all the others are not renormalizable. It can be also shown that the processes
with energy higher than $\Lambda_{\omega}^{(A)}$ become strong coupling, where $\Lambda_{\omega}^{(A)}$ is given by,
\bq
\lb{3.25}
\Lambda_{{\omega}}^{(A)}  = \frac{16}{81} \left(\frac{M_{pl}}{M_{A}}\right)^{3}M_{pl}\left|c_{\psi}\right|^{4},\; (M_{A} < M_{B}).
\eq
Therefore, when the fourth-order derivative terms dominate, the strong coupling problem still exists. This is expected, as power-counting tells us that the theory is
renormalizable only when $z \ge 3$ [cf. Eq.(\ref{1.1})]. Indeed, as to be shown
below, when the sixth order derivative terms dominate, the strong coupling problem does not exist any longer.

\subsubsection{$M_{A} \gtrsim M_{B}$}

In this case,  we have $M_{*} = M_{B}$, and   for  processes with $E \gtrsim M_{B}$,   Eq.(\ref{3.9}) reduces to
\bq
\lb{3.26}
S^{(2)} = \beta^{2}\int{dtd^{3}x\left(\dot\chi^{2}    -  \frac{1}{M_{B}^{4}} \chi\partial^{6}\chi\right)}.
\eq
Then,  by the transformations (\ref{3.12}) with  
\bq
\lb{3.27}
b_{2} = \frac{b_{1}^{1/3}}{M_{B}^{2/3}},\;\;\;
b_{3} = \frac{M_{B}}{\beta},
\eq
 we obtain,
\bq
\lb{3.28}
S^{(2)} =  \int{d\hat{t}d^{3}\hat{x}\left(\hat\chi^{*2}    - \hat\chi\hat\partial^{6}\hat\chi\right)},
\eq
while   the cubic action $S^{(3)}$ becomes,
  \bq
 \lb{3.10d}
 S^{(3)} = \frac{M_{B}}{\beta^{3}} \hat{S}^{(3)}.
   \eq
Eq.(\ref{3.28}) is invariant under the rescaling,
\bq
\lb{3.29}
\hat{t} \rightarrow b^{-3}\hat{t},\;\;\;
\hat{x}^{i} \rightarrow b^{-1}\hat{x}^{i},\;\;\;
\hat{\chi} \rightarrow \hat{\chi}.
\eq
Then, it can be shown that   the first five terms in Eqs.(\ref{3.10d}) and (\ref{3.10b}) are scaling-invariant, and so the last term. The terms of $\lambda_{6, 7, 8}$, on the other hand,
 scale, respectively, as $b^{-6},\;
b^{-4},\; b^{-2}$. Therefore, the first five  and the last terms now all become strictly renormalizable, while the   $\lambda_{6},\; \lambda_{7}$ and $\lambda_{8}$ terms become superrenormalizable
\cite{Pol}. To have these strictly renormalizable terms to be weakly coupling, we require their coefficients be less than unit,
\bq
\lb{3.30}
\frac{M_{*}}{\beta^{3}} \lambda_{n} < 1, \; (n = 1, ..., 5, 9).
\eq
For $\lambda \sim 1$ (or $|c_{\psi}| \sim 0$), we find that the above condition holds for
$$
M_{*} < \frac{2}{3} M_{pl}\left|c_{\psi}\right|.
$$
It can be shown that this condition holds identically, provided that $M_{*} < \Lambda_{\omega}$, that is,
\bq
\lb{3.32}
 M_{*} <  \left(\frac{\zeta}{c_1}\right)^{3/2}M_{pl}\left|c_{\psi}\right|^{5/2}.
\eq
[Recall $\Lambda_{\omega}$ is given by  Eq.(\ref{3.20}) and $M_{*} = M_{B}$.]
One can take $c_{1} \simeq M_{pl}$, but now a more reasonable choice is   $c_1 \simeq M_{*}$. Then, the condition (\ref{3.32}) becomes
 \bq
\lb{3.31}
M_{*} <   M_{pl}\left|c_{\psi}\right|^{1/2},\; (c_1 = M_{*}),
\eq
 which is much less restricted than the one of   $c_1 \simeq M_{pl}$. 
In addition, in order to have the sixth order derivative terms dominate, we must also require, 
\bq
\lb{3.33}
 M_{A} \gtrsim M_{*}.
\eq
Therefore, it is concluded that, {\em provided that  the conditions (\ref{3.32}) and (\ref{3.33}) hold,   the HMT  setup with any $\lambda$ \cite{HMT,Silva} does not have
 the strong coupling problem}.

\section{Conclusions}
\renewcommand{\theequation}{4.\arabic{equation}} \setcounter{equation}{0}

In this paper, we have studied the strong coupling problem of a scalar field in the framework of the HMT setup \cite{HMT}, with an arbitrary coupling constant $\lambda$, generalized recently by
da Silva \cite{Silva}. As shown previously  \cite{HW}, when the energy of a process is higher than   $\Lambda_{\omega}$, it becomes strong coupling. To avoid it, one can 
provoke the Blas-Pujolas-Sibiryakov (BPS) mechanism \cite{BPS}, in which a new energy scale $M_{*}$ is introduced, so that the sixth order derivative terms become important before the strong
coupling energy scale $\Lambda_{\omega}$ reaches. Once the high order derivative terms take over, the scaling behavior of the system is modified in such a way that all the nonrenornalizable terms
become either strictly renormalizable or supperrenormalizable, as shown explicitly in Sec. III.B.2. Whereby,  the strong coupling problem is resolved. 

It should be noted that, in order for the  mechanism to work, $\lambda$ cannot be exactly one, as one can see from Eq.(\ref{3.32}). In other words, the theory cannot reduce exactly to general
relativity  in the IR. However, since general relativity has achieved great success in low energies, $\lambda$ cannot be significantly different from one in the IR, in order for the theory to be consistent with  observations. 
As first noticed by BPS in their model without projectability condition, the most stringent constraints   come from the 
preferred frame effects due to Lorentz violation, which requires  \cite{BPS},
\bq
\lb{4.1}
|\lambda - 1| \lesssim 4\times 10^{-7},\;\;\; M_{*} \lesssim 10^{15} \; {\mbox{GeV}}.
\eq
In addition,  the timing of active galactic nuclei \cite{Timing} and gamma ray bursts \cite{GammaRay} requires
\bq
\lb{4.2}
M_{A} \gtrsim 10^{10} \sim  10^{11} \; {\mbox{GeV}}.
\eq
To obtain the constraint (\ref{4.1}), BPS used the results from the Einstein-aether theory, as these two theories  coincide in the IR  \cite{Jacob}. 

 In this paper, we have shown that the BPS mechanism  is also applicable to the HMT setup. However, it is not clear whether the condition (\ref{4.1}) is also applicable to the HMT setup,
 as  the effects due to Lorentz violation  in this setuop   have not been worked out, yet. On the other hand, 
the condition (\ref{4.2}) is  applicable, because this condition was obtained  from the dispersion relations, which are the same in both setups. 

In addition, the BPS mechanism cannot be applied to the  Sotiriou-Visser-Weinfurtner
generalization  with projectability condition \cite{SVW},  because  the condition $M_{*} < \Lambda_{\omega}$, together with the one that instability cannot occur within the age of the universe, 
requires fine-tuning, 
\bq
\lb{4.3}
|\lambda - 1| < 10^{-24},
\eq
as shown explicitly in \cite{WWb}. However, in the HMT setup with any $\lambda$, the Minkowski spacetime is stable \cite{HW}, so such a fine-tuning does not exist.

~\\{\bf Acknowledgements:}
The authors would like to thank Tony Padilla   for valuable discussions and comments.
AW  is supported in part by DOE Grant, DE-FG02-10ER41692 and NNSFC, 11075141; and 
QW $\&$  TZ  were supported in part by  NNSFC grant, 11047008.


\end{document}